\documentstyle[aps,twocolumn,prb,epsf]{revtex}
\begin{document}
\title{
The Kronig-Penney-Ising picture of the colossal magnetoresistance}
\author{N. Vandewalle$^{1}$, M. Ausloos$^{1}$, and
R. Cloots$^{2}$}
\address{$^{1}$SUPRAS, Institute of Physics, B5, University of Li$\grave e$ge,
B-4000 Li$\grave e$ge, Belgium\\
$^{2}$SUPRAS, Institute of Chemistry, B6, University of
Li$\grave e$ge, B-4000 Li$\grave e$ge, Belgium\\}
\date{\today}
\draft
\maketitle
\begin{abstract}
From general arguments, it is shown that a magnetic Kronig-Penney model
based on the thermodynamics of an Ising model can be used for describing
the Colossal Magnetoresistance (CMR) phenomenon. The model considers a
tunneling-like transmission process of hopping electrons through a dynamic
lattice characterized by evolving magnetic clusters. In this model,
correlations between the magnetic states are considered to be more relevant
than the lattice strain effects for obtaining the CMR features. Physical
arguments lead to the theoretical description of the intrinsic temperature
and field dependences of the CMR observed in typical manganite materials.
\end{abstract}

\narrowtext

It is often argued that the colossal magnetoresistance (CMR) is due to a
complicated interplay between electronic and lattice degrees of freedom
(polaron). In the following, we show that the magnetic degrees of freedom
contain the basic contribution to the field and temperature CMR behavior.

Let us recall the intriguing CMR phenomenon. The resistivity of perovskite
materials has an ``enormous" peak at some temperature $T_{M-I}$ considered
as related to a metal-insulator (M-I) transition. This transition is often
accompanied by a second-order ferromagnetic transition at a temperature
$T_c$. Under a magnetic field, the $\rho(T,H)$ appreciably decreases near
the temperature where $\rho(T,0)$ has its maximum. Many theories have been
proposed for CMR, but the exact temperature and field behaviors are not yet
fully satisfactory understood.

The CMR phenomenon in perovskite based materials is herebelow explained
through critical phenomenon behaviors and analytical laws for describing
the main observations. The model seems much more simple than previously
presented models. At this stage, it should not be requested that it solves
all known puzzles for all compounds with CMR. Simplifications are made
herebelow for presenting the approach. In fact, we present an alternative
view with respect to the most popular theories based on the magnetic polaron
idea \cite{polaron} including the Double Exchange Scenario (DES) \cite{des}.
Various features seem hard indeed to be put into this polaronic-DES framework,
e.g. at high temperature the small polaron picture with a few meV activation
energy due to lattice distortion \cite{distorsion} does not directly match to
the large polaron picture of the metallic state at low temperature, where a
band picture should emerge and hold below the magnetic
transition \cite{largepolaronband}. In fact, the question can be raised
whether there is a change in the mobility of carriers or in their number
(or both) or whether there is even a change in the band structure necessarily
implying some conductivity transition. For instance, it can be accepted that
the mean free path is small at high temperature, but at low temperature the
zero field resistivity dependence seems to be that of metallic
ferromagnets \cite{Mottferrom,liu}. Also transport features at fixed
magnetization $m$ indicate that for large or small magnetization values the
exponentially stretched dependence is markedly
different \cite{wisconsinpaper,gu,qli,eckstein}.
The role of lattice strain as shown on films grown on different substrates
does not seem to explain the findings, and should be supplemented by taking
into account magnetic domain like effects \cite{pennstatepaper}. This implies
that magnetic features should be reemphasized. The DES has been recently also
much criticized because it was thought that it could not explain the
qualitative difference in electrical conduction for the whole range of $x$
in one of the most often studied CMR materials, i.e. $La_{1-x}Ca_{x}MnO_3$.
However, it has been shown that a more correct treatment of the DES leads to
an acceptable view in the interesting doping range [$x=0.16 ,0.40$], i.e.
where the magnetic and conducting transitions are high (near room temperature).
Outside this interval, further work should be envisaged since the magnetic
structure has quite another periodicity, but the following concepts should
still hold as it will be easily seen.
In fact, three simple ingredients can be combined in a realistic way in order
to emphasize the magnetic degrees of freedom role, i.e. the Ising model for
the magnetic spins, the Drude formula for the electrical conductivity, and
some type of scattering for hopping electrons. By combining these basic
ingredients, we essentially take into account magnetic cluster effects, but
the more so their correlated fluctuations. The former scattering strength and
magnetic state life time are the only microscopic physical parameters.
Notice that

(i) Grains are usually pretty small, whence there is much grain boundary
scattering. We neglect such a (extrinsic) background term here, though it can
be easily inserted in the scattering strength if necessary.

(ii) The number of carriers is also kept field and temperature independent.
To take into account a density of states temperature (and even field)
dependence is a rather trivial generalization to be made, within a
self-consistent picture taking into account band and localized state carriers
(with possible spin states).

(iii) The temperature dependent lattice distortion \cite{distorsion} role for
hopping charge carriers is also neglected here, but can be included at a
later stage again in the definition of a temperature dependent lattice
parameter and mapped into an effective carrier mass or into an effective
localized spin coupling strength.

In so doing, we do not claim that the present model should give precise
quantitative values at this time because of the extremely limited number of
parameters which we are using. Nevertheless the theory will be in good
agreement with experimental data. Moreover, extensions seem easy in light of
the {\it paraphernalia} of solid state physics ideas and techniques. This
model is surely not the unique alternative to the magnetic polaron model.
However, this paper shows that the correlation between spins is the key
ingredient to be integrated in the understanding of the CMR phenomenon, seen
as a transport property in presence of magnetic states rather than a set of
such near equilibrium specific states controlled by some unknown exchange
interaction.
For the following developments, let us reduce the problem to a two
dimensional ($d=2$) case, allowing as a first approximation an in-plane
conduction like in thin films \cite{pennstatepaper}.
The Ising model \cite{onsager} on a $d=2$ square lattice is used for the
spins on the manganese sites assumed to represent the local magnetization of
the system. This simplified picture allows for a faster way of obtaining
the following results, but the spin-spin exchange interactions could be as
well of indirect origin as in DES without loss of generality. Simply we let
each lattice site $i$ contains a two-state spin $\sigma_i = \pm 1$. The
dimensionless Hamiltonian reads
\begin{equation}
E = - K \sum_{i,j}{\sigma_i \sigma_j - h \sum_i{\sigma_i}}
\end{equation}
where $K = J/kT$ is the dimensionless interaction between nearest neighboring
$i,j$ spins, and $h$ is the dimensionless magnetic field $H/kT$ orienting the
spins. The ($h=0,K>0$) case is a classical problem taught in classrooms
because it has a non-trivial phase transition as demonstrated by Onsager,
i.e. a logarithmic divergence of the specific heat near the reduced critical
temperature $K_c = - \frac{1}{2} \ln{(\sqrt{2}-1)}$. The Ising model implies
that there are droplets (clusters) of, e.g. $+1$ or $-1$ spins which nucleate,
grow, coalesce and disappear as a function of temperature \cite{dyn}. It is
well known \cite{onsager,stanley} that a ferro-paramagnetic transition takes
place exactly at $K_c$ for $h=0$ on such a lattice. In each grain or if the
intergrain coupling is adequate, clusters of respectively up and down spins
coexist and the average size $\xi$ of these clusters diverges at $K_c$
following $\xi \sim |K - K_c|^{-\nu}$ with $\nu=1$ \cite{stinchcombe}. The
other properties like $m(K,h)$ are not known exactly because the Ising model
in a field has not yet been solved.
In view of the partially covalent-ionic bonding in the plane, the quasi
localized carriers are supposed to be (spinless) electrons having a linear
hopping motion along the electrical field imposed across the lattice. The
Lorentz force is neglected here because of the rather short mean free path.
In the computer experiments, we launch electrons toward the right at random
from the left side of the ``sample". Each electron jumps to the right from a
site to the next nearest neighbor site at each time step  as follows. In a
so-called ``magnetic cluster", the electron hopping is free. When an electron
reaches a magnetic cluster wall, the electron is stopped with a probability
$(1-p)$ or transmitted with a probability $p$. By analogy with tunneling
effect, $p$ is assumed to be the exponential of some measure of the cluster
size $s$ ahead of the electron, i.e.
\begin{equation}
p = \exp{(-\gamma s)}
\end{equation}
where $\gamma$ is a dimensionless parameter which is like a potential barrier
strength of the cluster. There is no retention time upon a site nor phonon
nor magnetic drag, nor other type of scattering. At each time step the
magnetic structure is recalculated according to a Monte-Carlo procedure for
the Ising Hamiltonian. We count the carrier arrival time $\tau$ on the right
hand side of the ``sample". This time obviously depends on the sign
distribution fluctuations of the spins for a given $K$ and $h$ on the line
during the electron hopping.

Following the Drude formula, the resistivity is directly obtained from
\begin{equation}
\rho = {\tau \over L}
\end{equation}
where $L$ is the size of the lattice. At high temperature, when the spins
are completely disordered the resistivity is of course large; it is smaller
but not negligible at lower temperature; near the critical point $K_c$ for
$h=0$, the resistivity should become enormous: indeed the electron is a
little bit ``at a loss" because the spin fluctuations are huge and much
hamper the electron motion. A magnetic field stiffens the clusters (or
reduces the fluctuations). Therefore, the resistivity should be reduced
because the electron has a greater chance to find its way through. Thus, the
qualitative features of CMR are immediately found in this simple model.
Notice that this CMR version is somewhat like a temperature dependent
"magnetic Kronig-Penney model" in an electric field since each wall is a
potential barrier of which strength $\gamma$ is controlled by magnetic and
thermal conditions (Fig.1b), just like in disordered thin
films \cite{kpm,stoll}. The non-trivial (new) ingredient is that the
"barriers" are correlated and controlled by the thermodynamics of the Ising
model, in space and time.
In order to obtain a good numerical convergence, we have left the magnetic
system to reach a pseudo-steady-state before launching the ``electrons".
Lattices up to $256 \times 256$ were used. For conciseness, we fixed
arbitrary herein $\gamma=1$. The results do not change drastically with
$\gamma$. Clearly, at a later stage of investigations, the $\gamma s$ term
can be itself temperature and/or magnetization dependent for taking into
account the lattice strain. More complicated schemes taking into account
different spin channels can be also imagined within an effective medium
approximation. Here we consider that the spin density corresponds to the
case where the majority of polarized (up or down) spins is much larger than
the minority. The spinless approximation of the charge carrier is not even
a strong approximation. Indeed, it is clear that the true electron hopping
is depending on the availability of a neighboring state of similar nature.
Thus, the electron will be rather stopped in front of a wall after which the
spins states have the opposite sign to that of the incoming electron. This
is also in the DES spirit in fact.
Figure 2 presents the resistivity $\rho$ as a function of $K$. Both zero and
non-zero magnetic field cases are illustrated. The ferro-paramagnetic
transition at $K_c$ is indicated by the vertical line. As expected, a bump
is observed in $\rho$ below $K_c$ and an inflexion at $K_c$. Moreover, the
bump height decreases as $h$ increases. Such a bump can be viewed as the
signature of a percolation transition \cite{percobump} rather than a strict
metal-insulator transition. On both sides of $K_c$, the resistivity $\rho$
decreases exponentially. This shows that the experimentally observed
decreasing behavior in the high temperature phase might have nothing to do
with a semiconducting phase or a ``metal-insulator" transition as often
claimed, but rather to the number of available final state in the scattering.
The dimensionless excess resistivity defined by
\begin{equation}
\Delta \rho = {\rho(0) - \rho(h) \over \rho(0)}
\end{equation}
is shown as a function of $K$ in Figure 3. Data due to different values of
the magnetic field are shown. This quantity $\Delta \rho$ is found to be
independent of the length $L$ of the lattice. The qualititative features of
the CMR are well observed, i.e. (i) a peak at some intermediate temperature,
(ii) an increase of $\Delta \rho$ with $h$, and (iii) a shift of the peak
towards high temperatures and (iv) a wide transition region.
A non-trivial test of the model and theory is in order on Figure 4 which
presents the evolution of the maximum of $\Delta \rho$ as a function of $h$
in a semi-log plot. The value of $\Delta \rho$ for $K = K_c$ is also given.
A logarithmic increase of $\Delta \rho$ with $h$ is observed, i.e.
\begin{equation}
\Delta \rho \sim \ln{h}.
\end{equation}
Such a behavior can be found in $La_{1-x}Ca_xMnO_3$,
$La_{1-x} Mg_x MnO_3$, and $Pr_{1-x} Sr_x MnO_3$
compounds \cite{pennstatepaper,wagner} (Figure 5). The predicted logarithmic
behavior (straight lines) seems to hold quite well for moderate to high
magnetic fields. It is true that one should distinguish between low field,
moderate and high field CMR \cite{fontcuberta}. At lower field values (e.g.
$H<0.2$ T) inhomogeneities play a relevant
role \cite{pennstatepaper,fontcuberta}, an effect outside the present
investigation.
In addition, since CMR can be seen as related to a second-order critical
phenomenon, it should be possible to describe CMR with the help of scaling
arguments.
For $h=0$, the time for an electron to cross the system of size $L$ is
approximately given by
\begin{equation}
\tau = L + {L \over \xi} \sum_{i=1}^{+\infty} {{\left( 1- \exp{(-\gamma \xi)}
\right)}^i},
\end{equation}
for $\xi < L$ where $\xi$ is the characteristic size of the clusters and
$(1-\exp{(-\gamma s)})^i$ is the probability that the electron remains
blocked $i$ successive times on a magnetic wall before being transmitted to
the next site. Using the Drude formula, one has
\begin{equation}
\rho = 1 + {1 \over {\xi \ln{\left( 1- \exp{(-\gamma \xi)} \right)}}}.
\end{equation}
The above relationship neglects the correlations between wall fluctuations.
Nevertheless, the form of Eq.(8) explains the finite size effects, (Figure 2),
to be seen when $\xi \approx L$ and for small $L$ values. Close to the
ferro-paramagnetic transition, one can develop Eq.(8) and obtain
\begin{equation}
\rho \sim \exp{(\xi)} \sim \exp{\left({1 \over |K - K_c|}\right)}.
\end{equation}
The wall fluctuation correlations being in fact negligeable far away from
$K_c$, the above exponential scaling behavior seems to be correct on both
sides of $K_c$. A quantitative comparison to available data is not immediate
at low field ($H < 300mT$) since grain boundary background is very much
field dependent there. In such a regime because of inherent inhomogeneity
of (usually small) grains, the $\Delta \rho (h)$ dependence is indeed to be
known as non universal \cite{judith}.
In the case of large $h$ values, it has been argued elsewhere \cite{field}
that the in-plane correlation length $\xi$ near $K_c$ scales as
\begin{equation}
\xi \sim h^{- \nu /(b \delta)}
\end{equation}
where $b=1/8$ and $\delta=15$ for the $d=2$ Ising model whence the
$\Delta \rho \sim \ln h$ behavior should not be expected. Introducing the
latter scaling relation in Eq.(9) leads to a stretched exponential, a law
which qualitatively implies some apparent parallelism of $\Delta \rho (h)$
decaying curves with an amplitude being temperature dependent. This feature
can be understood as resulting from the non-negligable far away from $K_c$
cluster fluctuations and from some drift due to the magnetic field $h$.

It may be recalled that an analogous treatment to ours occured in the
pioneer theoretical work of Fisher and Langer \cite{fisher} using a $d=2$
Ising model instead of a true $d=3$ model for describing the experimental
results on the resistivity at ferromagnetic transitions. A divergence was
predicted at $T_c$ while experimentally an inflexion (only) was seen. This
paradoxical situation was found to be due to using a 2D rather than a 3D
Ising model. Extensions towards a better agreement to $d=3$ cases should
follow the same generalizing lines \cite{Mottferrom} in the future for our
CMR model.

\acknowledgments

NV is financially supported by the FNRS. A grant from FNRS/LOTTO allowed to
perform specific numerical work. MA acknowledges informal discussions with
I. Bozovic, F.R. Bridges, J. Fontcuberta, Qi Li, R.B. Stinchcombe, M.S.
Rzchowski and H. Vincent.

\newpage

\begin{figure}[htb]
\epsfxsize=8cm
\centerline{\epsffile{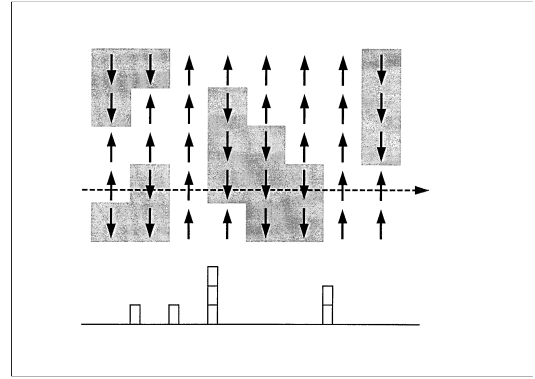} }
\caption{
Schematic illustration of the magnetotransport process as discussed in the
text: (a) The square lattice on which +1 and -1 magnetic domains are
distributed. On this lattice, one electron follows the linear motion
illustrated by the dashed arrow; (b) The barrier landscape viewed by this
electron at the time corresponding to the snapshot of (a).
}
\end{figure}

\begin{figure}[htb]
\epsfxsize=8cm
\centerline{\epsffile{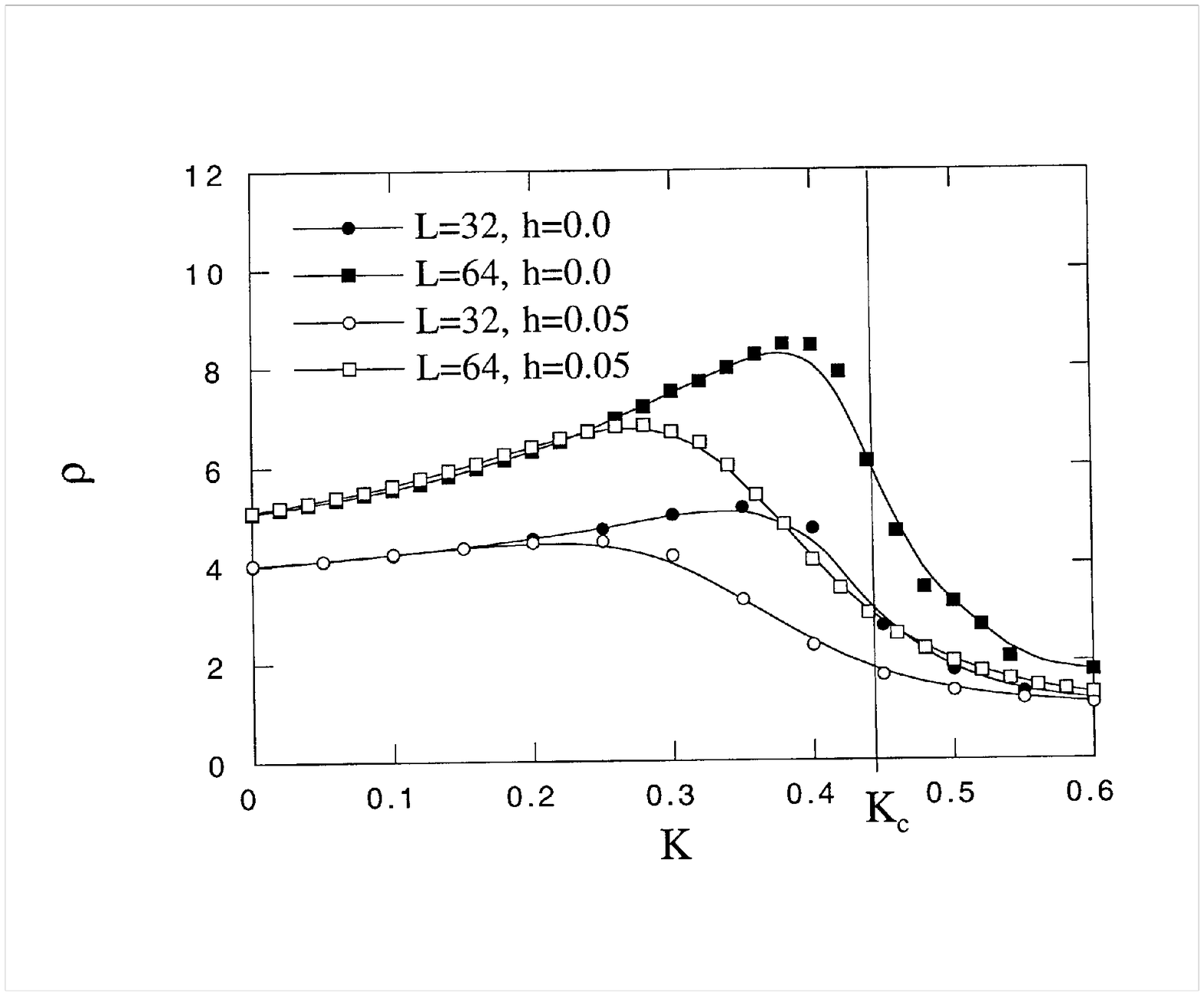} }
\caption{
Theoretical resistivity $\rho$ as a function of $K$. Two cases are shown:
$h=0$ and $h \ne 0$. Different lattice sizes are illustrated: $L=32$, $L=64$.
}
\end{figure}

\begin{figure}[htb]
\epsfxsize=8cm
\centerline{\epsffile{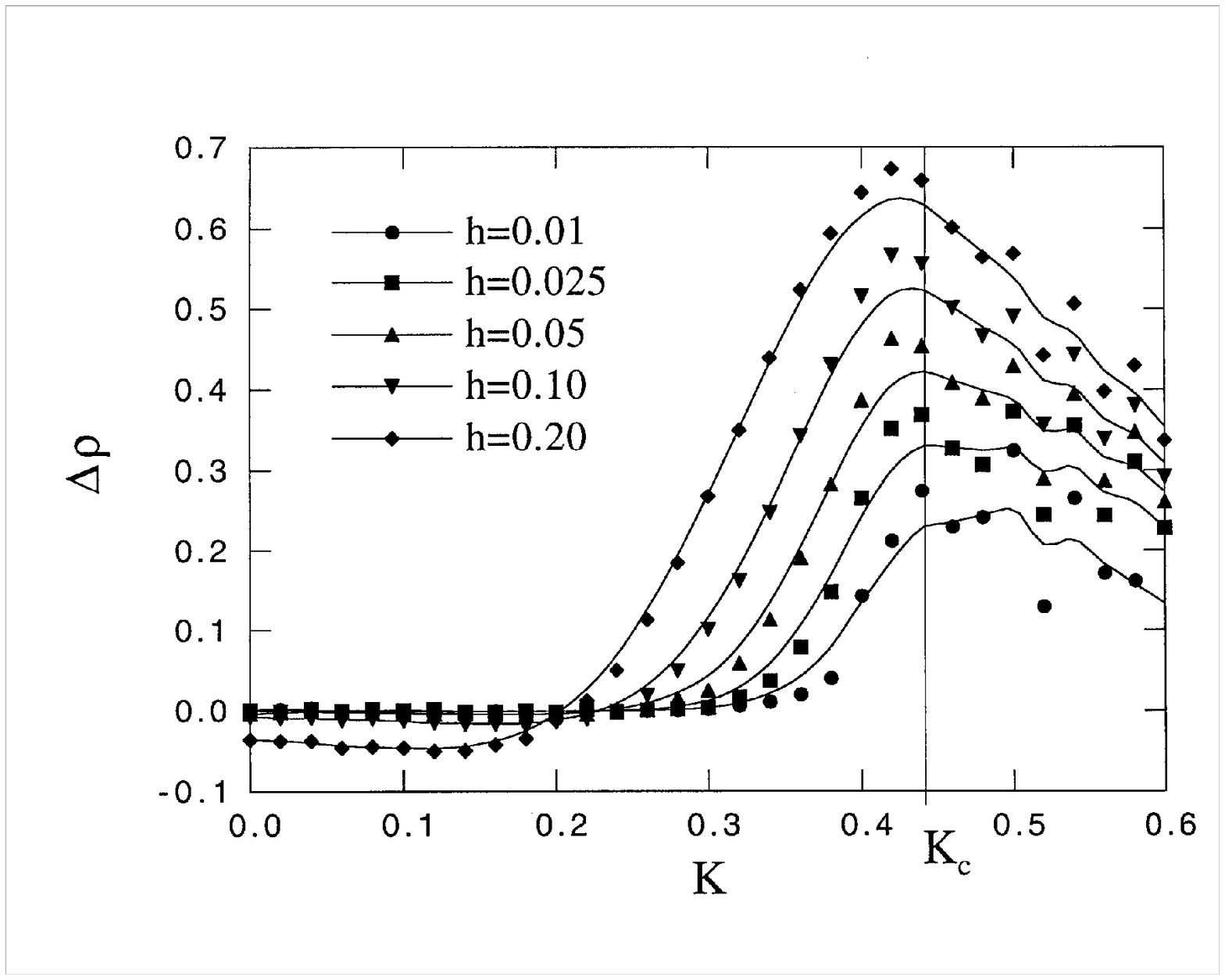} }
\caption{
Magnetoresistance $\Delta \rho$ as a function of $K$.
}
\end{figure}

\begin{figure}[htb]
\epsfxsize=8cm
\centerline{\epsffile{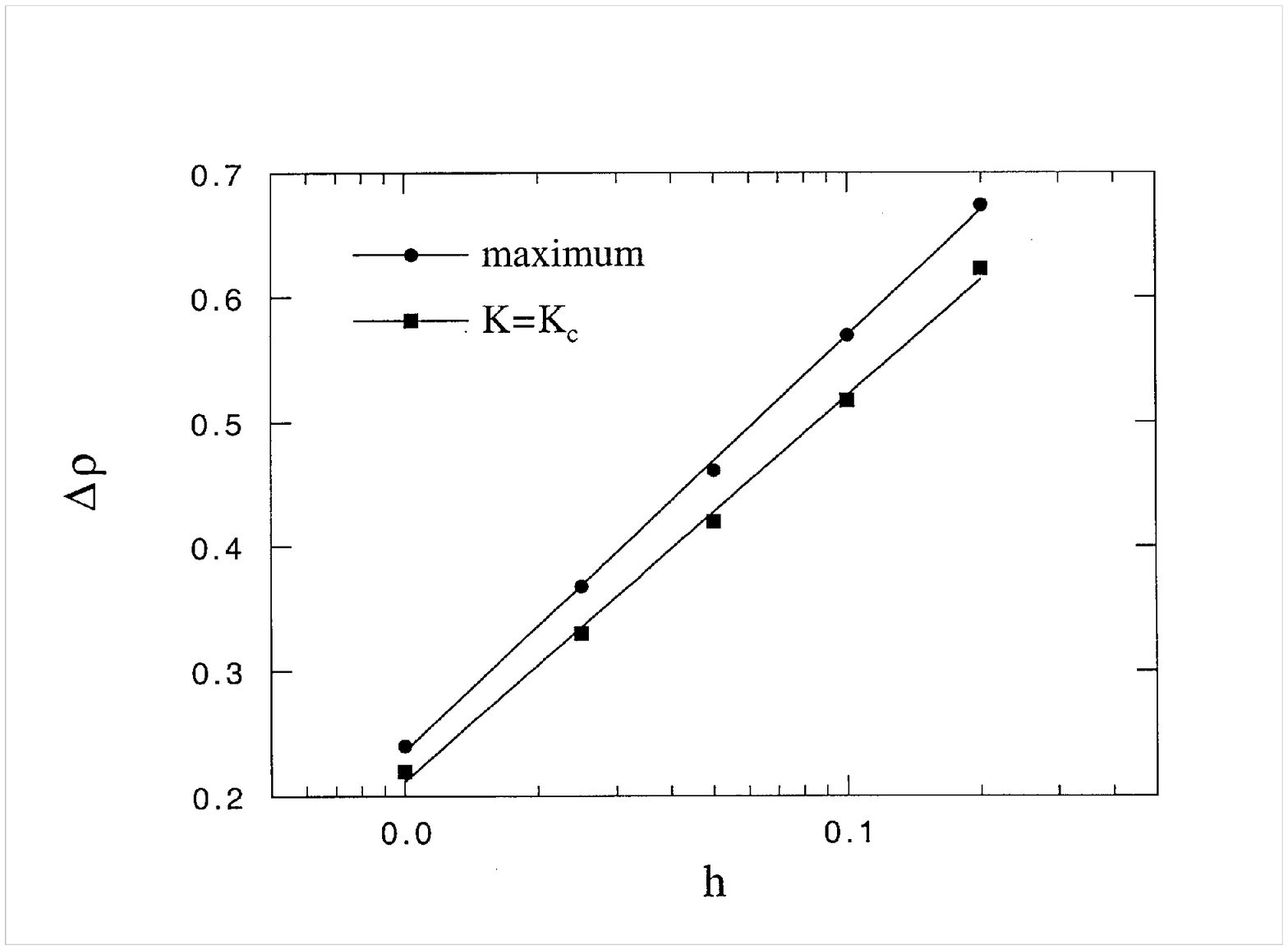} }
\caption{
Semi-log plot of the evolution of the maximum of the magnetoresistance
$\Delta \rho$ as well as $\Delta \rho$ for $K = K_c$ as a function of the
magnetic field $h$.
}
\end{figure}

\begin{figure}[htb]
\epsfxsize=8cm
\centerline{\epsffile{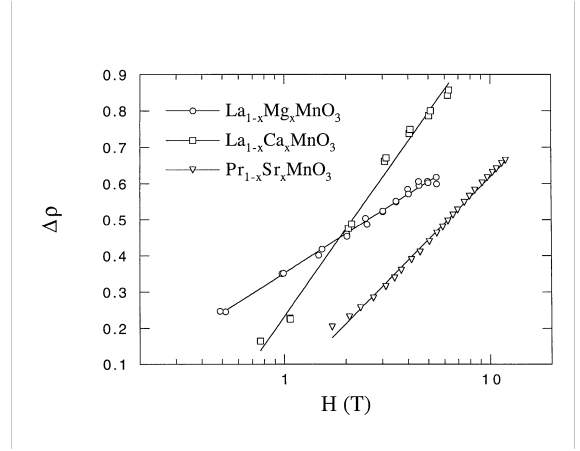} }
\caption{
Semi-log plot of the evolution of the magnetoresistance $\Delta \rho$ as a
function of the magnetic field for various compounds: $La_{1-x}Ca_xMnO_3$
from \cite{pennstatepaper}, $La_{1-x}Mg_xMnO_3$ from \cite{pennstatepaper}
and $Pr_{1-x}Sr_xMnO_3$ from \cite{wagner}. The logarithmic behavior
(straight line) predicted by the model (Eq.(6)) is clearly observed.}
\end{figure}
\end{document}